\documentclass[journal]{IEEEtran}
\usepackage{cite}

\ifCLASSINFOpdf
  \usepackage[pdftex]{graphicx}
  \graphicspath{{../pdf/}{../jpeg/}}
\else
\fi
\usepackage{subcaption}
\usepackage{graphicx}  

\usepackage[cmex10]{amsmath}
\usepackage{amsfonts} 
\usepackage{algorithmic}

\newtheorem{remark}{Remark}

\usepackage{array}
\usepackage{fixltx2e}

\usepackage{stfloats}
\usepackage{xcolor}
\usepackage{soul}
\soulregister\cite7
\soulregister\ref7 
\definecolor{Skyblue}{RGB}{135,206,235} 
\definecolor{Black}{RGB}{0, 0, 0}

\usepackage{stfloats}
\usepackage{diagbox}
\usepackage{multirow}

\usepackage{tabularx,booktabs}
\usepackage{makecell}
\usepackage{url}

\begin{document}
%
\title{Network Slicing for eMBB, URLLC, and mMTC: 
An Uplink Rate-Splitting Multiple Access Approach}
%
%
%

\author{{Yuanwen Liu,   
       Bruno~Clerckx,~\IEEEmembership{Fellow,~IEEE,}
      Petar~Popovski,~\IEEEmembership{Fellow,~IEEE,}}
      %
 \thanks{Yuanwen Liu is with the Department of Electrical and Electronic Engineering, Imperial College London, London SW7 2AZ, U.K. (e-mail: y.liu21@imperial.ac.uk).}
 \thanks{Bruno Clerckx is with the Department of Electrical and Electronic Engineering, Imperial College London, London SW7 2AZ, U.K. and with Silicon Austria Labs (SAL), Graz A-8010, Austria (e-mail:b.clerckx@imperial.ac.uk).}
\thanks{Petar Popovski is with the Department of Electronic Systems, Aalborg University, 9220 Aalborg, Denmark (e-mail: petarp@es.aau.dk).}}
\maketitle

\begin{abstract}
There are three generic services in 5G: enhanced mobile broadband (eMBB), ultra-reliable low-latency communications (URLLC), and massive machine-type communications (mMTC). To guarantee the performance of heterogeneous services, network slicing is proposed to allocate resources to different services. Network slicing is typically done in an orthogonal multiple access (OMA) fashion, which means different services are allocated non-interfering resources. However, as the number of users grows, OMA-based slicing is not always optimal, and a non-orthogonal scheme may achieve better performance. This work aims to analyse the performances of different slicing schemes in uplink, and a promising scheme based on rate-splitting multiple access (RSMA) is studied. RSMA can provide a more flexible decoding order and theoretically has the largest achievable rate region than OMA and non-orthogonal multiple access (NOMA) without time-sharing. Hence, RSMA has the potential to increase the rate of users requiring different services. In addition, it is not necessary to decode the two split streams of one user successively, so RSMA lets suitable users split messages and designs an appropriate decoding order depending on the service requirements. This work shows that for network slicing RSMA can outperform NOMA counterpart, and obtain significant gains over OMA in some regions.
\end{abstract}

\begin{IEEEkeywords}
eMBB, mMTC, network slicing, rate-splitting multiple access, URLLC
\end{IEEEkeywords}

%
\IEEEpeerreviewmaketitle

\section{Introduction}
%
%
%
%
\IEEEPARstart{5}{G} features three generic services: enhanced mobile broadband (eMBB), ultra-reliable and low-latency communications (URLLC), and massive machine-type communications (mMTC). These three services have various applications and requirements \cite{intro,8685766}. eMBB aims to provide extremely high data rates with high reliability, for applications such as augmented reality and remote presence. mMTC aims to provide connectivity for numerous devices which are cost and energy-constrained. The objective of URLLC is to provide a service featuring ultra-reliability and low latency while the rate is relatively low compared to mMTC. Some typical use cases are shown in Table \ref{tab:1} \cite{HUAWEI,5G_america,report,5G_america_mmtc}.
\begin{table}[]
\renewcommand{\arraystretch}{1.5}
\caption{Requirements of eMBB, URLLC and mMTC applications}
\label{tab:1}
\centering
\begin{tabular}{|*{5}{c|}}
\hline
\multicolumn{2}{|c|}{\diagbox{Applications}{Requirements}} & \thead{User experienced \\ data rates} & Latency & Reliability \\
\hline

\multirow{2}{*}{eMBB} &   Uplink file transfer & 50 Mbps & 16 s & 99.9\%\\
\cline{2-5}  

   & AR/VR & 50 Mbps & 10 ms & 99 \% \\

\hline
\multirow{2}{*}{URLLC} & Factory automation  & 10 Mbps & 10 ms & 99.99\% \\
\cline{2-5}
    & Motion control& 1 Mbps & 1 ms& 99.9999\% \\
\hline
\multirow{2}{*}{mMTC} & \thead{Sensing and factory \\ logistics} & 1 Mbps & 0.5 s & 95\% \\
\cline{2-5}
& Wearables & 2-5 Mbps& Relaxed & N/A\\
\hline
\end{tabular}
\end{table}

Network slicing is a promising technique to satisfy different service requirements \cite{8004168,7926923}. This technique allocates communication resources to the users requiring different services to guarantee their varied requirements. Usually, network slicing is done by OMA-based slicing. However, as the number of users grows, a more flexible and efficient scheme may be explored, and these three services have different requirements, so network slicing may leverage these differences to get better performance. We note that this work is limited to the wireless part of the network slicing, and sometimes it is referred to as spectrum slicing \cite{e23060686}.

NOMA-based network slicing scheme has the potential to improve the performance in the uplink, and it was studied in \cite{8476595,9448942,9448974}. In \cite{8476595}, a NOMA-based slicing scheme was explored, and this work showed that NOMA-based slicing is more suitable than OMA-based slicing in some scenarios. In \cite{9448942}, network slicing for eMBB and URLLC coexistence was analysed. There was one eMBB user and multiple URLLC users in this scenario. When URLLC users had better channel conditions than the eMBB user, NOMA-based slicing always outperformed OMA-slicing. While when eMBB users had better channel conditions, NOMA-based slicing only had better performance when the eMBB rate was relatively high. In \cite{9448974}, the eMBB and mMTC coexistence scenario was analysed. There was one eMBB user and multiple mMTC users, and the arrival process of mMTC traffic followed Poisson distribution. This paper analysed the achievable pairs of eMBB rate and mMTC arrival rate. According to the results, NOMA-based slicing can achieve the pairs that OMA-based slicing cannot, and vice versa. As the number of receive antennas at the base station (BS) increased, NOMA-based slicing benefited more from this increase than OMA-based slicing. In \cite{8647460}, the performance trade-offs between the eMBB rate and the URLLC latency were investigated in NOMA-based slicing with both puncturing and successive interference cancellation (SIC) methods.

\IEEEpubidadjcol

Applying NOMA-based slicing in downlink has also been studied and showed lower power consumption and more effective utilization. In \cite{saggese2021noma, 9268188,8643428,9011578,8932425,9364885,chen2021urllc}, eMBB and URLLC coexistence was studied. In \cite{saggese2021noma}, NOMA-based slicing was employed to minimize the transmission power and it had lower power consumption than OMA-based slicing. \cite{9268188,8643428,9011578,8932425} showed that NOMA-based slicing can leverage resources more effectively, and it has promising performance when resources allocation and scheduling are designed properly, and \cite{9364885} proposed a learning-based approach for resource allocation. In \cite{chen2021urllc}, a scheduling algorithm for eMBB and URLLC coexistence in multiple-input multiple-output (MIMO) NOMA systems was studied, and it brought a higher spectrum utilization than the OMA counterpart. 

RSMA has drawn a lot of attention recently for both downlink and uplink communications \cite{Mao2022RateSplittingMA,https://doi.org/10.48550/arxiv.2205.02548}. In the downlink, the message of one or multiple user(s) is split into a common part and a private part, and all the common
parts are encoded jointly into one common stream while the private parts are encoded independently into private streams. Each user can then reconstruct the original message from the decoded common stream and decoded private stream. By splitting messages at the transmitter, RSMA has been shown to unify and outperform NOMA and space division multiple access (SDMA) \cite{article,8907421}. RSMA was shown to be not only more spectrally efficient but also more energy efficient \cite{8491100}. \cite{9663192} showed that RSMA is a powerful technique in multi-user MIMO scenarios. In \cite{xu2021ratesplitting}, RSMA was shown to obtain the same performance as SDMA and NOMA with shorter blocklength, therefore opening the door to low latency applications. \cite{9491092} showed that RSMA is more robust than SDMA to the degrading effects of mobility and latency in CSI acquisition. With these advantages, RSMA is a competitive candidate, and \cite{9451194,9348672} have shown that RSMA is a very promising technique for future networks.

RSMA has promising performance not only in the downlink but also in the uplink. For uplink RSMA, a user splits its message into two streams and transmits a superposed message to the BS \cite{Mao2022RateSplittingMA,485709}. How the power is split between the two streams is a critical aspect because it dictates the performance. We use the splitting power fraction to refer to the power fraction allocated to the first stream without loss of generality. Since the number of streams increases, RSMA can provide a more flexible decoding order at the BS. By wisely deciding the decoding order and adjusting the splitting power fraction, RSMA can achieve all boundary points of the capacity region with SIC without the need for time-sharing \cite{485709,9257190}, improve the user fairness and outage performance \cite{9064705,liu2022rate}, and have the potential to simplify the implementation for uplink by avoiding the need for user pairing \cite{8171078}. \cite{Liu2021RateSM} demonstrated that RSMA could increase connectivity and reliability in a semi-grant-free transmission scenario, which are significant expectations for future networks. A RSMA-based slicing scheme for eMBB and URLLC coexistence was proposed in \cite{9643016}, and this work showed that RSMA could achieve a larger rate region than NOMA-based and OMA-based slicing when the power splitting factor is properly configured. 

In this paper, we explore the performance of RSMA-based slicing\footnote{RSMA-based slicing is called RSMA for simplicity in the rest of the paper.} in eMBB and URLLC coexistence and eMBB and mMTC coexistence. For eMBB and URLLC coexistence, the contributions of this paper are summarised as follows:
\begin{itemize}
\item This paper investigates eMBB and URLLC coexistence with RSMA and analyses the performances in different channel conditions. The achievable rate regions and the relation between the splitting power fraction and the sum-rate of URLLC users are shown. Although \cite{9643016} analysed the eMBB and URLLC coexistence when channel conditions of URLLC users were better than eMBB, and showed that RSMA always outperformed OMA-based slicing\footnote{OMA-based slicing is called OMA for simplicity in the rest of the paper.} and NOMA-based slicing\footnote{NOMA-based slicing is called NOMA for simplicity in the rest of the paper.} when users of the same services shared the resource in a non-orthogonal fashion, it did not analyse the opposite scenario when an eMBB user experiences a better channel than URLLC users. In the previous scenario, it would be easier for RSMA to exploit flexible decoding order; while in the later one, sometimes RSMA may boil down to NOMA. Interestingly, the later scenario shows when RSMA can outperform OMA and reveals a suitable scenario for RSMA. Our results show that RSMA can always achieve a larger rate region compared to NOMA, while it outperforms OMA in some scenarios. Adjusting splitting power fraction appropriately can improve the performance, and the lower the interference from the eMBB user, the higher the improvements RSMA can make. 

\item This paper gives a detailed analysis of the suitable scenario for RSMA. The main superiority of RSMA in uplink comes from the flexible decoding order since the split streams do not have to be decoded successively. Hence, for the user with the lower rate, one of its split streams can be decoded after the streams of other users, so it has lower interference and achieves a higher rate; for the user with a higher rate, its stream can be decoded early to deal with more interference. Thus, all the users can achieve a higher rate. In this way, a higher decoding order flexibility leads to higher achievable rate pairs. This is the reason why RSMA has better performance when eMBB interference is not that high. 
The achievable rate regions of NOMA and RSMA in eMBB and URLLC coexistence are also simulated to demonstrate the suitable scenario of RSMA.
\end{itemize} 

For eMBB and mMTC coexistence, the contributions are summarised below: 
\begin{itemize}
\item This paper proposes a RSMA scheme for eMBB and mMTC coexistence, and to the best of our knowledge, RSMA has not been applied to this scenario before. In this scheme, the message of the eMBB user is split into two streams. By adjusting the splitting power fraction of eMBB and scheduling the decoding order, the signal-to-interference-plus-noise ratio (SINR) of mMTC devices is also adjusted, since the first stream of eMBB message can be cancelled before decoding some mMTC devices. Consequently, more mMTC devices can have sufficient SINR to satisfy the target rate compared to NOMA, which means a higher arrival rate of mMTC can be supported using RSMA for a given reliability requirement.

\item This paper analyses the performance of eMBB and mMTC coexistence with RSMA. The achievable pairs of the eMBB rate and the arrival rate of mMTC and the relation between the splitting power fraction of eMBB and the arrival rate of mMTC are shown. According to the simulation results, RSMA can achieve rate pairs non-achievable by NOMA and OMA, and RSMA is shown to leverage the difference between reliability requirements of different services better compared to NOMA.

\end{itemize}

The rest of this paper is organised as follows. Section \ref{sec:2} gives the introduction to the system model. Section \ref{sec:3} analyses the network slicing performances of eMBB and URLLC coexistence. Section \ref{sec:4} introduces eMBB and mMTC coexistence. Section \ref{sec:5} presents the numerical results, and Section \ref{sec:6} is the conclusions.

\textit{Notations}: $\mathbb{C}$ represents the set of complex numbers. $\mathbb{E}\left[ \cdot \right]$ refers to the statistical expectation. $\mathcal{CN}(\delta,\sigma^2)$ represents a complex Gaussian distribution with mean $\delta$ and variance $\sigma^2$. $\mathrm{Poisson}(\lambda)$ represents a Poisson distribution with mean $\lambda$. $\Gamma (s,x)$ represents the upper incomplete gamma function which is defined as  $\Gamma (s,x)=\int_x^{\infty} t^{s-1} e^{-t} \ dt$. $Q(x)$ represents Gaussian $Q$ function which is $Q(x)=\int_x^{\infty}{\frac{1}{\sqrt{2\pi}} \mathrm{exp}\left( \frac{-t^2}{2}\right) dt}$.

\section{System Model}\label{sec:2}
\begin{figure}[]
\centering
\includegraphics[scale=0.4]{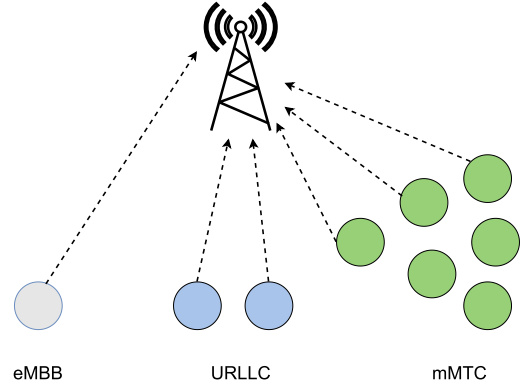}
\caption{The considered scenario that devices from three services transmit to a common BS in the uplink.}\label{fig:diagram}
\end{figure}

\begin{figure}[]
\centering
\includegraphics[scale=0.55]{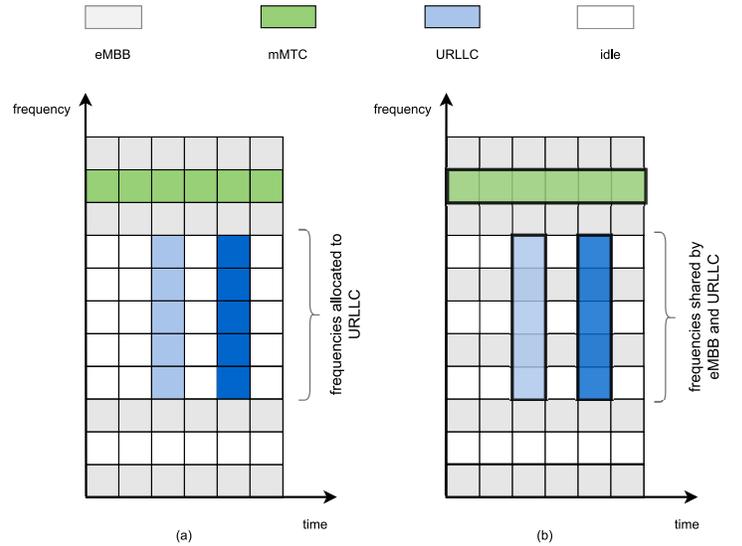}
\caption{Illustration of network slicing. (a) is the OMA scenario, and (b) is the general non-orthogonal slicing scenario, i.e., NOMA and RSMA. mMTC and eMBB, eMBB and URLLC can share the resources. The overlap of URLLC transmission is represented by dark blue, i.e., multiple URLLC users transmit simultaneously.}\label{fig:sling_illustration}
\end{figure}

This section introduces the model of 5G services individually, namely eMBB, URLLC and mMTC, and a general signal model of RSMA. Different services can share the same radio resources when they communicate to a common BS as shown in Fig. \ref{fig:diagram}. We consider that there are $F$ frequency channels indexed by $f\in\{1,...F\}$, and $S$ time slots indexed by $s\in\{1,...S\}$, as shown in Fig. \ref{fig:sling_illustration}. Fig. \ref{fig:sling_illustration}a is the illustration of OMA. Each service occupies individual resources without overlap, but users of the same service can share the resources, i.e. one URLLC user can be allocated an individual resource denoted by light blue, and multiple URLLC users can also be allocated to the same resources represented by dark blue. Fig. \ref{fig:sling_illustration}b shows the general non-orthogonal slicing scenario, which includes NOMA and RSMA. Different services can share the same resource, and the sharing part is denoted by the overlap.

We assume that eMBB users are allocated to single or multiple radio resources at one or several given frequencies, while URLLC users can occupy several radio frequencies and a single mini-slot because of the stringent latency requirement\footnote{Besides latency requirement, high reliability is also an important requirement of URLLC and diversity is critical to achieve it \cite{8362303}. Therefore, transmitting with multiple frequencies would be more reasonable to increase diversity. In this way, increasing diversity compensates for fades in wireless channels \cite{8815506,8456628} and guarantees reliability.}, and mMTC users are allocated a specified frequency resource. The numbers of URLLC users and active mMTC devices are $n_U$ and $n_M$, respectively. The arrival process of mMTC follows a Poisson distribution, $n_M\sim \text{Poisson}(\lambda_M)$, where $\lambda_M$ is the mean value and referred to as the arrival rate of mMTC users.

As in \cite{8476595}, we assume that the radio access and competition among eMBB devices have been solved prior to the considered time slot, and there is only one eMBB device in each time slot. Each frequency channel $f$ is assumed to be within the time-and frequency-coherence interval of the wireless channel, so the wireless channel coefficients are seen as constant. The channel coefficients of eMBB user, URLLC user $u$ and mMTC user $m$ at frequency $f$ are denoted by $h_{B,f} \in \mathbb{C}$, $h_{U,u,f}\in \mathbb{C}$ and $h_{M,m,f}\in \mathbb{C}$, respectively, where $u=1,2,...n_U$ and $m=1,2,...n_M$, and they fade independently. The channels are considered as Rayleigh fading channels, i.e. $h_{B,f} \sim \mathcal{CN}(0,\Gamma_B)$, $h_{U,u,f} \sim \mathcal{CN}(0,\Gamma_U)$, and $h_{M,m,f} \sim \mathcal{CN}(0,\Gamma_M)$, where $\Gamma_B$, $\Gamma_U$ and $\Gamma_M$ are the average channel gains for eMBB, URLLC and mMTC users, respectively. The channel gains of the users of three services in radio frequency $f$ are denoted by $G_{B,f}=|h_{B,f}|^2$, $G_{U,u,f}=|h_{U,u,f}|^2$, and $G_{M,m,f}=|h_{M,m,f}|^2$.
Without loss of generality, the power of the noise at BS is normalized to one. Thus, the received power equals SNR. The received signal at BS at time slot $s$ and frequency $f$ can be represented as
\begin{equation}
\begin{aligned}
    y_{s,f}=&h_{B,f} x_{B,s,f}+\sum^{n_U}_{u=1} h_{U,u,f} x_{U,u,s,f}\\
    &+\sum^{n_M}_{m=1} h_{M,m,f} x_{M,m,s,f}+z_{s,f},
\end{aligned}
\end{equation}
where $x_{B,s,f}$, $x_{U,u,s,f}$, $x_{M,m,s,f}$ and $z_{s,f}$ are signals transmitted by eMBB user, URLLC user $u$, mMTC user $m$ and noise at the frequency $f$ and time slot $s$, respectively.

We assume that the eMBB device has perfect channel state information (CSI), while URLLC and mMTC do not have CSI, and BS also has perfect CSI as in \cite{8476595}. Thus, the eMBB device can adjust the transmission power according to the CSI, while for URLLC and mMTC devices the transmission power is fixed to 1.
The reliability requirements of eMBB, URLLC and mMTC are $\epsilon_B$, $\epsilon_U$ and $\epsilon_M$, respectively. Let $E_B$, $E_U$, $E_M$ denote the events that the eMBB user is in outage, the event URLLC users are in outage and the event mMTC users are in outage, respectively. Thus, the outage probabilities should satisfy these reliability requirements, i.e. Pr$(E_B) \leq \epsilon_B$, Pr$(E_U) \leq \epsilon_U$ and Pr$(E_M) \leq \epsilon_M$.

In the following subsections, models of eMBB, URLLC and mMTC will be introduced, and at this stage, they are considered individually without interference from other services. Then, a general uplink rate-splitting multiple access model will be briefly introduced, and the detailed uplink RSMA models for different coexistence scenarios will be introduced in Section \ref{sec:3} and Section \ref{sec:4}.

\subsection{eMBB}
Since the eMBB device and BS are assumed to have perfect CSI, the eMBB device can adjust the transmission power $P_B(G_{B,f})$ according to the instantaneous channel gains to achieve a predefined SNR value. Thus, the objective is that the eMBB device transmits at the largest rate $r_{B,f}$ at frequency $f$ and the outage probability is lower than reliability requirements $\epsilon_B$ under a long-term average power constraint, so the problem can be formulated as

\begin{equation}
\begin{aligned}
\max \ ~ &r_{B,f} \\
\mathrm{s.t.} \ \ ~ ~ &\text{Pr}\left[ \log_2 \left( 1+G_{B,f}P_B \left(G_{B,f} \right) \right) <r_{B,f} \right] \leq \epsilon_B \\
            &\mathbb{E}\left[ P_B \left(G_{B,f} \right) \right]=1.
\end{aligned}
\end{equation}
The optimal solution to this problem is given by truncated power inversion \cite{771147}, which means that the eMBB device chooses a transmission power that is inversely proportional to the channel gain $G_{B,f}$ if it is above a given threshold $G_{B,f}^{min}$. The probability that the eMBB transmits is
\begin{equation}
    a_B=\text{Pr}[G_{B,f}\geq G_{B,f}^{min}]=e^{-G_{B,f}^{min}/\Gamma_B}.
\end{equation}
Without interference from other services, the only outage source is that the eMBB user does not transmit due to the insufficient SNR level, so the event that the eMBB user is in outage is actually the event that the eMBB user is inactive, i.e. not transmitting. Thus, the error probability is 
\begin{equation}
    \text{Pr}(E_B)=1-a_B.
\end{equation}
Impose the reliability requirement $\epsilon_B$
\begin{equation}
    \text{Pr}(E_B)=\epsilon_B,
\end{equation}
so we can obtain the value of threshold SNR
\begin{equation}
    G_{B,f}^{min}=\Gamma_B \ln \left( \frac{1}{1-\epsilon_B} \right).
\end{equation}
With the power-inversion-scheme, the transmission power is chosen as a function of the instantaneous channel gain $G_{B,f}$ as
   
\begin{equation}
     P_B(G_{B,f})=
    \begin{cases}
       \frac{G_{B,f}^{tar}}{G_{B,f}} & \text{if $G_{B,f}\geq G_{B,f}^{min}$} \\
        0 \  & \text{if $G_{B,f}< G_{B,f}^{min}$}\\
    \end{cases},
\end{equation}
where $G_{B,f}^{tar}$ is the target SNR. Imposing the average power constraint to be normalized to one,

\begin{equation}
    1=\mathbb{E}[P_B(G_{B,f})]= \int_{G_{B,f}^{min}}^{\infty} \frac{1}{\Gamma_B}e^{-x/\Gamma_B}P_B(x)\,dx, 
\end{equation}
and the target SNR can be obtained as
\begin{equation}
    G_{B,f}^{tar}=\frac{\Gamma_B}{\Gamma \left( 0,\frac{G_{B,f}^{min}}{\Gamma_B} \right)}.
\end{equation}
Thus, the largest transmission rate of eMBB can be obtained as 
\begin{equation}
    r_{B,f}=\log_2(1+G_{B,f}^{tar}) \ \mathrm{(bits/s/Hz)}.
\end{equation}
From this result, we know that $r_{B,f}$ only depends on $\Gamma_B$ and $\epsilon_B$, so for each frequency channel the largest transmission rate is the same, and the index $f$ can be omitted. This largest transmission $r_B$ is referred to as $r_B^{orth}$,
\begin{equation}\label{r_B}
    r_B^{orth}=\log_2(1+G_{B,f}^{tar}) \ \mathrm{(bits/s/Hz)}.
\end{equation}

\subsection{URLLC}
URLLC devices transmit messages across $F_U$ frequency resources, where $F_U \leq F$. We assume that URLLC devices do not have CSI, and the transmission power is fixed to 1. BS has the CSI of URLLC users and uses SIC to decode the messages. In the SIC procedure, BS receives the superposition of URLLC signals, successively detects the received data symbols, and then subtracts them from the superposition of signals. URLLC users are intermittent and each user is active with the probability $a_U$, so the number of active URLLC users is random in a given spanning time resource with multiple frequency resources, which we refer to as a frequency strip. This is a generalization of URLLC setup in \cite{8476595}, which assumes a single URLLC user. Here we consider the worst case that there are always $n_U$ URLLC users transmitting simultaneously in the given frequency strip, so the outage probability in practice should be lower than the estimation which will be given then. Let $G_{U,u,f}$ denote the channel gain of user $u$ in frequency channel $f$, and the user decoded after user $u$ is denoted by $j>u$, so the SINR of user $u$ in frequency $f$ 
\begin{equation}
    \sigma_{U,u,f}=\frac{G_{U,u,f}}{1+\sum^{n_U}_{j>u}G_{U,j,f}}.
\end{equation}
Thus the outage probability is
\begin{equation}\label{URLLC}
    \text{Pr}(E_U)=\frac{1}{n_U}\sum^{n_U}_{u=1} \text{Pr} \left( \frac{1}{F_U}\sum_{f=1}^{F_U}\log_2 \left(1+\sigma_{U,u,f} \right) <r_U\right) ,
\end{equation}
where $r_U$ is the target rate of URLLC.

The decoding order is decided by the sum of the mutual information across $F_U$ frequency resources \cite{9448942}. The users are successively decoded, and BS picks the device with the largest value of the sum of mutual information $I_{sum}$ to decode at each step, i.e., for user $u$ it is
\begin{equation}
    I_{sum}=\frac{1}{F_U}\sum^{F_U}_{f=1} \log_2 \left( 1+\frac{G_{U,u,f}}{1+\sum^{n_U}_{j>u}G_{U,j,f}} \right).
\end{equation}
Thus, we can impose the reliability requirement $\text{Pr}(E_U)=\epsilon_U$ and obtain the maximum $r_U$, and the sum-rate of URLLC users is
\begin{equation}\label{URLLC_sum}
    r_U^{sum}=n_U r_U  \ \mathrm{(bits/s/Hz)}.
\end{equation}
Increasing $F_U$ can enhance the frequency diversity, so it is possible to obtain a higher $r_U$ while satisfying the reliability requirement $\epsilon_U$ with larger $F_U$.

\subsection{mMTC}

The arrival process of mMTC devices is random, so we assume that it follows a Poisson distribution with the arrival rate $\lambda_M$, and the rate of mMTC devices is fixed at $r_M$. CSI is unknown for mMTC, and the transmission power is fixed to 1. Since mMTC devices are allocated a specified frequency resource, the index $f$ is omitted. 

BS uses SIC to decode mMTC devices. The decoding order is decided by the channel gains of users. Let $m$ denote the mMTC device with the $m$-th largest channel gain and the channel gains are sorted in descending manner $G_{1}\geq G_{2}\geq...G_{n_M}$, so the decoding order is $1, 2...n_M$. Since there is no interference from other services, for a random user $m_0$, the SINR is
\begin{equation}\label{mMTC_SINR}
    \sigma_{m_0}=\frac{G_{m_0}}{1+\sum^{n_M}_{m=m_0+1}G_{m}}.
\end{equation}
If $\log_2(1+\sigma_{m_0})\geq r_M$ holds, this mMTC device can be decoded successfully and then cancelled. Otherwise, it cannot be decoded and the procedure terminates. Let $D_M$ denote the number of successfully decoded devices, and the error probability measures the ratio of successfully decoded devices $D_M$ to all active devices $n_M$. Thus the error probability is 
\begin{equation}
    \text{Pr}(E_M)=1-\frac{\mathbb{E}[D_M]}{\lambda_M},
\end{equation}
where $\mathbb{E}[D_M]$ is the expectation of $D_M$, and $\lambda_M$ can represent the mean value of $n_M$. Thus, the problem can be formulated as 
\begin{equation}\label{mMTC_arrival}
    \lambda^{orth}_M(r_M)=\text{max}\{\lambda_M : \text{Pr}(E_M)\leq\epsilon_M\},
\end{equation}
and the value of $\lambda^{orth}_M(r_M)$ can be computed by Monte Carlo method.

\subsection{Signal Model of Rate-Splitting in the Uplink}
Rate-splitting method for the multiple-access channel was first introduced in \cite{485709}. The main idea of rate-splitting is to divide a message of one user into two parts, which is equivalent to adding a virtual user, and in this way provide a more flexible decoding order. For a general RSMA network slicing model in uplink, we assume that there is one eMBB user, two URLLC users and $n_M$ mMTC users. This setup can effectively capture the essence of the problem, because in a given cell the number of connected mMTC devices can be very large compared to the number of connected eMBB and URLLC users in each time slot. Since the arrival process of mMTC is random and mMTC devices are energy and cost-constrained, we assume that mMTC does not split the messages and the transmission power of the signal is fixed to 1. Hence, eMBB and URLLC users can have their message split, in contrast to mMTC users. Thus, the received signal of BS at a given time slot $s$ and frequency $f$ is ($s$ and $f$ are omitted for simplicity)

\begin{equation}
\begin{aligned}
    y= & h_{B} \sum_{i=1}^2 \sqrt{P_{B,i}} x_{B,i} 
  +\sum^{2}_{u=1} \sum_{i=1}^2 h_{U,u} \sqrt{P_{U,u,i}} x_{U,u,i}\\ 
           &+\sum^{n_M}_{m=1} h_{M,m} x_{M,m}+z,
\end{aligned}
\end{equation}
where $x_{B,i}$ and $x_{U,u,i}$ are split messages of eMBB user and URLLC user $u$, respectively, and $P_{B,i}$, ${P_{U,u,i}}$ are the power allocated for $x_{B,i}$ and $x_{U,u,i}$, respectively. For eMBB and URLLC, the power constraints are $\mathbb{E}\left[ \sum_{i=1}^2 \sqrt{P_{B,i}} \right]= 1$ and $\sum_{i=1}^2 \sqrt{P_{U,u,i}}=1$, respectively. 

We note that RSMA can unify NOMA and OMA in a single multiple access framework. When the transmission power is all allocated to one of the split streams and the other stream is not allocated power, RSMA boils down to NOMA; while when all the transmission power is turned off except one service, RSMA becomes OMA. Thus, OMA and NOMA can be contained in RSMA if full power control is applicable for all the users. However, since URLLC and mMTC do not have the CSI and the transmission power is fixed, they do not have the full power control, so for network slicing, OMA is not a subset of RSMA.

Although RSMA increases the number of messages, the decoding order is more flexible and it can achieve any boundary points of the rate region as shown in \cite{485709}. Other methods that achieve the same region are NOMA with time-sharing and joint encoding/decoding \cite{Mao2022RateSplittingMA}. However, implementing time-sharing requires multiple time slots \cite{9257190} and induces communication overhead  \cite{Mao2022RateSplittingMA}. A joint encoding/decoding approach is not practical to implement due to the high decoding complexity of random codes \cite{485709,Mao2022RateSplittingMA}. Since URLLC has stringent latency requirements and mMTC is aimed at cost and power-constrained devices, NOMA with time-sharing and joint encoding/decoding can be problematic in network slicing, so they will not be considered here.

\section{Slicing for eMBB and URLLC}\label{sec:3}
In this section, the coexistence of eMBB and URLLC will be discussed, and mMTC traffic will not be considered here. Since the radio access and competition
among eMBB devices are assumed to be solved, there is only one eMBB device and two URLLC devices in each time slot. 

\subsection{OMA for eMBB and URLLC}

In the OMA scenario, we assume that the eMBB device and URLLC devices are allocated individual frequency resources. Let $F$ denote the total number of the frequency resources, and URLLC devices are allocated $F_U$ frequency resources.

The performance of this system is specified in terms of $(r_B^{sum}, r_U^{sum})$ of eMBB sum-rate $r_B^{sum}$ and sum-rate of URLLC $r_U^{sum}$ while reliability requirements $(\epsilon_B,  \epsilon_U)$ are satisfied. The sum-rate of eMBB is obtained as 
\begin{equation}
    r_B^{sum}=(F-F_U)r_B^{orth}  \ \mathrm{(bits/s/Hz)},
\end{equation}
where $r_B^{orth}$ is obtained in (\ref{r_B}). The rate of URLLC devices can be computed from (\ref{URLLC}) while satisfying $\text{Pr}(E_U)\leq\epsilon_U$ , and the sum-rate can be computed from (\ref{URLLC_sum}).

\subsection{NOMA for eMBB and URLLC}

In the NOMA scenario, eMBB and URLLC devices share all $F$ frequency resources. Here BS uses SIC to decode eMBB and URLLC users, and it is shown in \cite{8476595} that the SIC method outperforms the puncturing method. Like the scenario mentioned in Section \ref{sec:2}-B, the worst case that there is always URLLC traffic when eMBB is transmitting is considered. Since URLLC devices are sensitive to latency, URLLC devices should be decoded prior to the eMBB device. The reliability requirement of URLLC is much more stringent than eMBB. If decoding eMBB first and this eMBB device fails, decoding URLLC users will fail. Thus, decoding eMBB first will require eMBB to satisfy the same reliability requirement as URLLC, and it is actually not possible for eMBB to satisfy it, so decoding eMBB first can deteriorate the reliability of URLLC. Hence, the decoding priority should be given to URLLC because it can leverage the reliability requirements difference. In addition, eMBB aims to provide the service with a high data rate, so decoding eMBB after URLLC will let eMBB users experience less interference and achieve a higher rate. Thus, the procedure is that URLLC devices are always decoded first. If it is successful, then an eMBB device will be decoded. Otherwise, the procedure terminates, because the interference from URLLC devices cannot be cancelled.

Since there is always interference from the eMBB device to URLLC devices, it would be better to choose a smaller eMBB target SNR to get better overall performance, i.e.

\begin{equation}\label{GB_tar_constraint}
    G_{B,f}^{tar}\leq \frac{\Gamma_B}{\Gamma \left( 0,\frac{G_{B,f}^{min}}{\Gamma_B} \right) },
\end{equation}
so the achievable rate for eMBB is
\begin{equation}\label{eMBB_rate}
r_{B,f}=\log_2 \left(1+G_{B,f}^{tar} \right) \ \mathrm{(bits/s/Hz)},
\end{equation}
and the sum-rate is
\begin{equation}\label{embb_sum_rate}
r_B^{sum}= \sum_{f=1}^{F} r_{B,f}.
\end{equation}

Given the rate of eMBB $r_{B,f}$, the objective is to find the highest achievable rate $r_U$ while reliability requirements are satisfied. The error probabilities of eMBB and URLLC are calculated individually. For eMBB, the error probability can be computed by the Monte Carlo method, which is
\begin{equation}\label{eMBB_error}
    \text{Pr}(E_B)=1-\mathbb{E}[D_B],
\end{equation}
where $D_B$ is the number of successfully decoded eMBB devices. 

Then the error probability of URLLC is formulated. The decoding order of URLLC can be decided by the same method as in Section II-B, and assuming it is $1,2,...n_U$, so the SINR of the URLLC device is 
\begin{equation}\label{urllc_sinr}
    \sigma_{U,u,f}=\frac{G_{U,u,f}}{1+\sum^{n_U}_{j>u}G_{U,j,f}+\delta_f G_{B,f}^{tar}} ,
\end{equation}
where $\delta_f$ is an independent Bernoulli random variable with parameter $a_B$ which is the active probability of the eMBB device. Since the reliability requirement of the eMBB user is also stringent, i.e. $10^{-3}$, it means that the probability that the eMBB device is inactive is extremely low. Thus, we consider the most stringent scenario that eMBB is always active, which means $\delta_f=1$ and there is always interference from the eMBB device. Thus, the error probability of URLLC can be presented by (\ref{URLLC}), where $\sigma_{U,u,f}$ is obtained from (\ref{urllc_sinr}) and $F_U=F$. Then, impose $\epsilon_U$, $\text{Pr}(E_U)\leq \epsilon_U$, and $r_U$ can be computed by the Monte Carlo method, and the sum-rate is $r_U^{sum}=n_U r_U  \ \mathrm{(bits/s/Hz)}$.

\subsection{RSMA for eMBB and URLLC}

Since URLLC devices are very sensitive to latency and have high reliability requirements, i.e. decoding of URLLC should not depend on decoding of other services, they should be decoded first, so rate-splitting is only used for URLLC devices 
\footnote{Actually, from an information theory perspective, all users but one (e.g. eMBB user and one of the two URLLC users, or the two URLLC users) split their messages to achieve the capacity region. However, in network slicing, latency and reliability should also be considered. Hence, due to similar reasons as in Section \ref{sec:3}-B, URLLC will always be decoded before eMBB and RSMA is only applied to URLLC users.}. 
Here we consider the setup in Section \ref{sec:2}-D with one eMBB user and two URLLC users and eMBB is always decoded after URLLC. If any URLLC message cannot be decoded, the procedure terminates and the eMBB message is lost. 

The eMBB part is the same as in Section \ref{sec:3}-B. $G_B^{tar}$ is chosen by the same method as in (\ref{GB_tar_constraint}), and the rate can be obtained from (\ref{eMBB_rate}) and (\ref{embb_sum_rate}). The objective is to find the highest achievable rate pairs ($r_U^{sum}$,$r_B^{sum}$) while satisfying the reliability requirements of eMBB and URLLC. The error probability of eMBB can be obtained from (\ref{eMBB_error}). The decoding order of URLLC is the same as the one in Section \ref{sec:2}-B. Let URLLC user 1 denote the device decoded first, and URLLC user 2 denote the other 
\footnote{Here we work with two assumptions on URLLC: (1) the instantaneous SNR is unknown and (2) the status of other URLLC user is unknown. Thus, BS will decide which user splits the message. Before exchanging data, the connection between URLLC users and BS needs to be set up. During this setup process, BS can obtain the CSI of URLLC users, and then decide which user will split the message and send the information of splitting message to this user.}.
The message from URLLC user 1 is divided into two parts, and the power factor for the first part is $\beta$, $\beta \in [0,1]$. The received signal at BS at frequency $f$ is
\begin{equation}
\begin{aligned}
    y=&h_{B,f} x_{B,f}+\sqrt{\beta} h_{U,1,f} x_{U,1,1,f}+\sqrt{1-\beta} h_{U,1,f} x_{U,1,2,f}
     &+h_{U,2,f} x_{U,2,f}+z,
\end{aligned}
\end{equation}
where $x_{U,1,1,f}$ and $x_{U,1,2,f}$ are the signals from the first and second part of the message of URLLC user 1, respectively. Without loss of generality, we assume that $ x_{U,1,1,f}$ is decoded first. The decoding order is set to $ x_{U,1,1,f}$, $x_{U,2,f}$ and $x_{U,1,2,f}$, since this order can obtain all the boundary points of rate region by adjusting $\beta$. Thus, the SINR of $x_{U,1,1,f}$  is
\begin{equation}
    \sigma_{U,1,1,f}=\frac{\beta G_{U,1,f}}{1+(1-\beta)G_{U,1,f}+G_{U,2,f}+G_B^{tar}},
\end{equation}
the SINR of $\mathbf{x}_{U,2,f}$ is
\begin{equation}\label{SINR_2}
    \sigma_{U,2,f}=\frac{G_{U,2,f}}{1+(1-\beta)G_{U,1,f}+G_B^{tar}},
\end{equation}
and the SINR of $\mathbf{x}_{U_{1,2}}$ is 
\begin{equation}
    \sigma_{U,1,2,f}=\frac{(1-\beta) G_{U,1,f}}{1+G_B^{tar}}.
\end{equation}
Then we get the corresponding rate
\begin{equation}
    r_{U,1,1,f}=\log_2(1+\sigma_{U,1,1,f}) \ \mathrm{(bits/s/Hz)},
\end{equation}
\begin{equation}
    r_{U,1,2,f}=\log_2(1+\sigma_{U,1,2,f})\ \mathrm{(bits/s/Hz)},
\end{equation}
\begin{equation}\label{R_2}
    r_{U,2,f}=\log_2(1+\sigma_{U,2,f}) \ \mathrm{(bits/s/Hz)},
\end{equation}
and the sum-rate of URLLC user 1 is 
\begin{equation}\label{rs_user_1}
    r_{U,1}=\frac{1}{F} \sum^{F}_{f=1}\left( r_{U,1,1,f}+r_{U,1,2,f} \right) \ \mathrm{(bits/s/Hz)}.
\end{equation}
Similarly, the rate of URLLC user 2 is
\begin{equation}\label{rs_user_2}
    r_{U,2}=\frac{1}{F} \sum^{F}_{f=1}r_{U,2,f} \ \ \mathrm{(bits/s/Hz)}. 
\end{equation}
The error probability is formulated as 
\begin{equation}\label{rs_error_probability}
    \text{Pr}(E_U)=\frac{1}{2}\sum_{u=1}^{2}\text{Pr}(r_{U,u}<r_U),
\end{equation}
and impose $\epsilon_U$, $\text{Pr}(E_U)\leq\epsilon_U$, and $r_U$ can be computed by Monte Carlo method, and $r_U^{sum}$ can be obtained from (\ref{URLLC_sum}).

\begin{figure}[]
\centering
\includegraphics[scale=0.6]{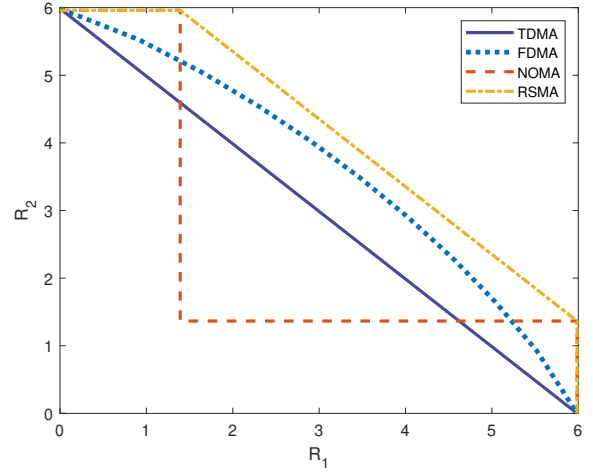}
\caption{An illustration of rate regions of TDMA, FDMA, NOMA and RSMA for a generic two-user case. The average channel gain of two users is $20$ dB.}\label{fig:illustration}
\end{figure}

\begin{remark}
The scenario with two URLLC users can also give insight into a more general scenario with $n_U$ URLLC users. First, we consider a generic two-URLLC-user case, and we do not consider NOMA with time-sharing for the reasons mentioned in \ref{sec:2}-D. We assume that user 1 splits the message into two streams, and user 2 does not split. If BS decodes the split streams of user 1 not successively, this decoding order can achieve any point on the diagonal line of the rate region. 
Fig. \ref{fig:illustration} is an illustration of rate regions of time-division multiple access (TDMA), frequency-division multiple access (FDMA), NOMA and RSMA for a generic two-user case, and the average channel gain of both users is $20$ dB. With the nonsuccessive decoding order, RSMA can achieve the points on the yellow diagonal line. Similarly, in URLLC and eMBB coexistence, RSMA could achieve all the boundary points of rate region by adjusting $\beta$. 
When there are two users, one user splits the message and these split streams are not decoded successively, this enables achieving any point on the diagonal line of the rate region by adjusting $\beta$. Thus, intuitively, when the number of users is $n_U$, these users can be ordered by mutual information in descending manner, and the order is $1, 2, ..., n_U$. Let the first $n_U-1$ users split messages and decode these split streams not successively, so this 
decoding order should also bring freedom to obtain any boundary points
of the rate region, i.e. the decoding order could be $x_{U,1,1}, x_{U,2,1},...,x_{U,n_U-1,1}, x_{U,n_U}, x_{U,n_U-1,2},..., x_{U,2,2}, x_{U,1,2}$.

\end{remark}

\section{Slicing for eMBB and mMTC}\label{sec:4}
In this section, the coexistence of eMBB and mMTC will be discussed. The arrival process of mMTC follows Poisson distribution as in Section II-C, and URLLC traffic will not be considered. Similarly, there is one eMBB device and multiple mMTC devices, because we assume that the radio access and competition
among eMBB devices are solved. Since only a single frequency channel is considered, the frequency indices $f$ are omitted in this section.

\subsection{OMA for eMBB and mMTC}

In the OMA scenario, eMBB and mMTC users share the resources in a time-sharing manner. Let $\alpha, \alpha \in [0,1]$, denote the time faction allocated to eMBB users, and the time allocated to mMTC users is $1-\alpha$. Thus, the achievable rate of eMBB is
\begin{equation}
    r_B=\alpha r_B^{orth}\ \mathrm{(bits/s/Hz)},
\end{equation}
where $r_B^{orth}$ is obtained in (\ref{r_B}). The arrival rate $\lambda_M$ can be computed by a method similar to that in Section II-C. It is equivalent to the scenario in Section II-C with mMTC rate at $\frac{r_M}{1-\alpha}$, so 
\begin{equation}
    \lambda_M=\lambda_M^{orth}\left(\frac{r_M}{1-\alpha}\right).
\end{equation}
Then, $\lambda_M$ can be obtained from (\ref{mMTC_arrival}).

\subsection{NOMA for eMBB and mMTC}

In the NOMA scenario, eMBB and mMTC devices share the same frequency resources at the same time. The decoding procedure is the same as in \cite{8476595}. Trying to decode the eMBB message before mMTC messages may be an option since the high rate requirement of eMBB can cause high interference when decoding mMTC messages. However, this is not the optimal option, because the error probability of mMTC is actually measuring the ratio of successfully decoded mMTC devices to all active mMTC devices instead of the decoding probability for each device. There may be some mMTC devices with very high channel gains, so it would be better to decode them before decoding the eMBB device.

Based on the analysis above, \cite{8476595} gave the exact procedure. The mMTC devices are ordered in a descending manner, and BS tries to decode mMTC devices one by one. If an mMTC device cannot be decoded, BS tries to decode the eMBB device. If this eMBB device is successfully decoded, then BS continues decoding mMTC devices. Otherwise, the decoding procedure terminates, because eMBB interference cannot be cancelled.

For the eMBB device, similar to the eMBB and URLLC coexistence scenario, it would be better to choose a lower target SNR for eMBB $G_B^{tar}$ which satisfies (\ref{GB_tar_constraint}), so the rate of eMBB device is (\ref{eMBB_rate}). Because the error of decoding the eMBB device is caused by the interference from mMTC devices and inactive state due to the insufficient SNR level, we must set a higher active probability for the eMBB device to satisfy the reliability requirement $\epsilon_B$. For example, if there is not any other interference, like the scenario in Section II-A, $a_B=1-\epsilon_B$, i.e. $\epsilon_B=10^{-3}$, $a_B=0.999$, while for eMBB and mMTC coexistence scenario, the active probability $a_B$ should satisfy $a_B>0.999$. Here we assume that $a_B=1$, which means there is always eMBB interference for mMTC devices, and the error of eMBB is only caused by decoding error. Thus, the SINR of an mMTC device decoded before the eMBB device is
\begin{equation}
    \sigma_{m_0}=\frac{G_{m_0}}{1+\sum_{m=m_0+1}^{n_M}G_{m}+G_B^{tar}}.
\end{equation}
If $\log_2(1+\sigma_{m_0}) \geq r_M$, this mMTC device can be decoded. Otherwise, it failed and BS will try to decode the eMBB device. The SINR of the eMBB device is
\begin{equation}
    \sigma_B=\frac{G_B^{tar}}{1+\sum_{m=m_0}^{n_M}G_{m}}.
\end{equation}
If $\log_2(1+\sigma_B)\geq r_B$, the eMBB device can be decoded, and then BS will try to decode other mMTC devices as in Section II-C. Otherwise, it fails and the procedure terminates.

Let $D_M$ and $D_B$ denote the number of decoded mMTC devices and eMBB device, respectively. The error probability of mMTC is the ratio of expectation of decoded devices to all active devices, $1-\mathbb{E}[D_M]/\lambda_M$, and the error probability of eMBB device is $1-\mathbb{E}[D_B]$. In order to find the achievable pairs $(r_B, \lambda_M)$, the problem can be formulated as
\begin{equation}
\begin{aligned}
    \lambda_M^{NOMA}(r_B)={\rm max} \{  \lambda_M & \geq 0: \exists G_B^{tar} \  \\ {\rm s.t.} \ & 
     \mathbb{E}[D_M]/\lambda_M\geq1-\epsilon_M \\ &
    {\rm and} \ \mathbb{E}[D_B] \geq 1-\epsilon_B\}.
\end{aligned}
\end{equation}
For a giving $r_B$, $\lambda_M^{NOMA}$ can be computed by Monte Carlo method.

\subsection{RSMA for eMBB and mMTC}

According to the discussion in NOMA for eMBB and mMTC, the decoding procedure can terminate when neither mMTC device nor eMBB device can be decoded. This implies that mMTC device and eMBB do not have sufficient SINR to achieve the target rate, which is caused by the relatively high interference from each other. Then the decoding procedure terminates because it is 'lodged'. 

To mitigate this effect, we apply RSMA in the eMBB and mMTC coexistence scenario. The eMBB message is split into two streams, so BS can decode one stream at a relatively low rate, and then continue decoding mMTC devices. Since part of the interference from eMBB can be cancelled, the decoding procedure avoids getting ‘lodged’ to some extent. The whole decoding procedure is described below. The decoding order of the mMTC devices is defined according to the descending order of their channel gains. Firstly, BS tries to decode mMTC devices. If a mMTC device cannot be decoded, BS tries to decode the first stream of eMBB at a relatively low rate $r_{B_1}$, and then continues decoding mMTC devices. Similarly, if another mMTC device cannot be decoded, BS tries to decode the second stream of eMBB at rate $r_{B_2}$. If $r_{B_1}+r_{B_2}<r_B$, the eMBB device failed, and the decoding procedure terminates. Otherwise, BS continues decoding mMTC devices as in Section II-C. Note that $r_{B_1}$ is a non-negative value, so it is always possible to decode the first stream with a rate not higher than $r_{B_1}$, so we assume that BS could always decode the first stream regardless of whether the second stream can be decoded successfully. Similar to Section IV-B, the active probability of eMBB is $a_B=1$, and the power factor for the first part of the eMBB message is $\beta$, $\beta \in [0,1]$. The received signal is
\begin{equation}
     y=\sqrt{\beta}h_{B} x_{B_{1}}+\sqrt{1-\beta}h_{B} x_{B_{2}}+
     \sum_{m=1}^{n_M}h_m x_m
     +z,
\end{equation}
where $x_{B_{1}}$ and  $x_{B_{2}}$ are the signals from the first and second streams of the eMBB message, respectively, and $h_m$ is the channel coefficient of mMTC user $m$. The target SNR of eMBB is still $G_B^{tar}$, and it satisfies (\ref{GB_tar_constraint}). We assume that $x_{B_{1}}$ is always decoded first, and its SINR at BS is 
\begin{equation}
    \sigma_{B_1}=\frac{\beta G_B^{tar}}{1+(1-\beta)G_B^{tar}+\sum_{m=m_1}^{n_M}G_m},
\end{equation}
where $m_1$ represents the index of the mMTC device which will be decoded after $x_{B_{1}}$, and the rate is
\begin{equation}
    r_{B_1}=\log_2(1+\sigma_{B_1}) \ \mathrm{(bits/s/Hz)}.
\end{equation}
The SINR of $x_{B_{2}}$ is
\begin{equation}
    \sigma_{B_2}=\frac{(1-\beta) G_B^{tar}}{1+\sum_{m=m_2}^{n_M}G_m},
\end{equation}
where $m_2$ represents the number of the mMTC device which will be decoded after $x_{B_{2}}$ , and the rate is
\begin{equation}
    r_{B_2}=\log_2(1+\sigma_{B_2}) \ \mathrm{(bits/s/Hz)}.
\end{equation}
For mMTC devices decoded before all eMBB streams, the SINR is
\begin{equation}
    \sigma_{m_0}=\frac{G_{m_0}}{1+G_B^{tar}+\sum_{m=m_0+1}^{n_M}G_m}.
\end{equation}
For mMTC devices decoded between $x_{B_{1}}$ and $x_{B_{2}}$, the SINR is
\begin{equation}
    \sigma_{m_0}=\frac{G_{m_0}}{1+(1-\beta)G_B^{tar}+\sum_{m=m_0+1}^{n_M}G_m}.
\end{equation}
For mMTC devices decoded after all eMBB streams, the SINR is as same as (\ref{mMTC_SINR}).

Let $D_M$ and $D_B$ denote the number of decoded mMTC devices and eMBB device, respectively. In this scenario, $D_B=1$ holds only when $r_{B_1}+r_{B_2}\geq r_B$ holds. Otherwise, $D_B=0$. For mMTC devices, they can be decoded when $\log_2(1+\sigma_{m_0})\geq r_M$ holds. Similarly, in order to find the achievable pairs $(r_B, \lambda_M)$, the problem can be formulated as
\begin{equation}
\begin{aligned}
    \lambda_M^{RSMA}(r_B)={\rm max} \{ \lambda_M & \geq 0: \exists G_B^{tar}\ {\rm and} \ \beta \ \\
    {\rm s.t.} \ &   
     \mathbb{E}[D_M]/\lambda_M\geq1-\epsilon_M \\&
    {\rm and} \ \mathbb{E}[D_B] \geq 1-\epsilon_B\}.
\end{aligned}
\end{equation}
$\lambda_M^{RSMA}$ can be computed by the Monte Carlo method.

\section{Numerical Results}\label{sec:5}
This section presents the simulation results of eMBB and URLLC coexistence and eMBB and mMTC coexistence. For eMBB and URLLC coexistence, the achievable rate pairs of eMBB and URLLC and the relation between splitting power fraction and sum-rate of URLLC in different channel conditions are shown and analysed. The rate region of two URLLC users in eMBB and URLLC coexistence is simulated to give a clearer analysis of what is a suitable scenario for RSMA. For eMBB and mMTC coexistence, the achievable pairs of eMBB rate and mMTC arrival rate and the relation between splitting power fraction and mMTC arrival rate are presented and analysed. The suitable situations for RSMA are also discussed in this section.

\subsection{eMBB and URLLC Coexistence}
\begin{figure}[]
\centering
\includegraphics[scale=0.6]{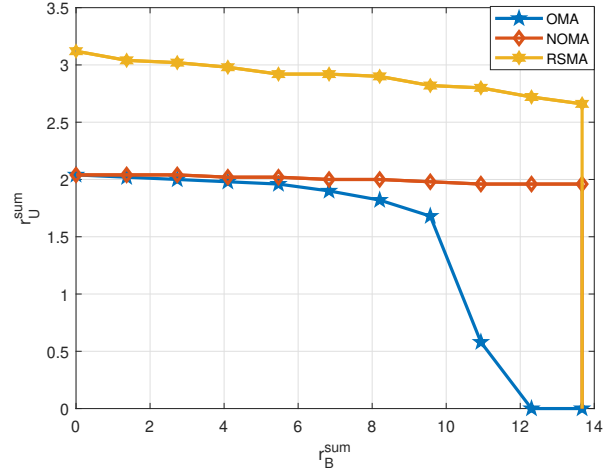}
\caption{Rate region $(r_B^{sum}, r_U^{sum})$, where $r_B^{sum}$, $r_U^{sum}$ are the sum-rate of eMBB and URLLC, respectively. $\Gamma_B=10$\ dB, $\Gamma_U=20$\ dB, $F=10$, $\epsilon_B=10^{-3}$, $\epsilon_U=10^{-5}$.}\label{fig:embb_urllc_leq}
\end{figure}

\begin{figure}[]
\centering
\includegraphics[scale=0.6]{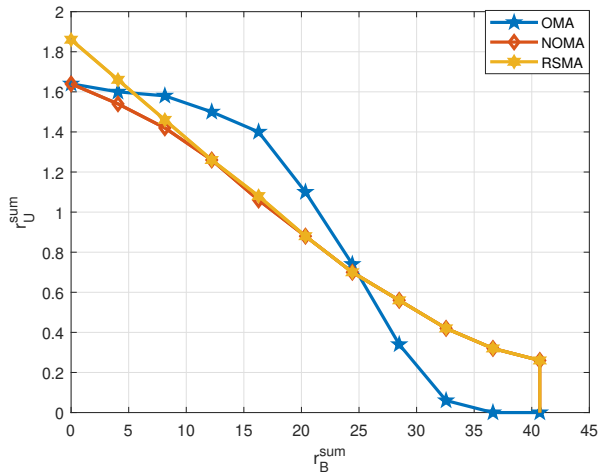}
\caption{Rate region $(r_B^{sum}, r_U^{sum})$, where $r_B^{sum}$, $r_U^{sum}$ are the sum-rate of eMBB and URLLC, respectively. $\Gamma_B=20$\ dB, $\Gamma_U=10$\ dB, $F=10$, $\epsilon_B=10^{-3}$, $\epsilon_U=10^{-5}$.}\label{fig:embb_urllc_geq}
\end{figure}

The simulation results when $n_U=2$ are shown. Rate regions $(r_B^{sum}, r_U^{sum})$ are presented in Fig. \ref{fig:embb_urllc_leq} and Fig. \ref{fig:embb_urllc_geq}. There is one eMBB user and two URLLC users, and the total number of frequency resources is $F=10$. Note that to make the figures more intuitive these rate regions are presented as enclosed, and each achievable rate pair is denoted by a marker.

In Fig. \ref{fig:embb_urllc_leq}, $\Gamma_B=10$ \ dB, $\Gamma_U=20$ \ dB, $\epsilon_B=10^{-3}$, $\epsilon_U=10^{-5}$. This result shows that NOMA always outperforms OMA, and RSMA outperforms these two schemes. In this scenario, $\Gamma_U>\Gamma_B$, non-orthogonal slicing schemes can leverage the difference between channel gains of URLLC and eMBB to perform SIC, so NOMA and RSMA outperform OMA, especially at higher rate of eMBB, and RSMA can always achieve a larger rate region than NOMA. The gain of RSMA comes from the decoding order flexibility. By allocating power to split streams appropriately, both URLLC users can achieve a higher $r_U$ and then a higher sum-rate $r_U^{sum}$ can be obtained. We can imagine a simple situation to get the intuition. Assuming that the channel gains of URLLC user 1 and URLLC user 2 are $21$\ dB and $19$\ dB, respectively. The target SNR of eMBB is assumed to $10$\ dB. If NOMA is applied and the decoding order is URLLC user 1, URLLC user 2 and eMBB, we can easily calculate the rate for URLLC user 1 and URLLC user 2 are $1.26\ \mathrm{(bits/s/Hz)}$ and $3.04\ \mathrm{(bits/s/Hz)}$. Obviously, $r_U$ should not be higher than $1.26\ \mathrm{(bits/s/Hz)}$ if we want both the two users to be decoded. If RSMA is applied and URLLC user 1 splits its message into two streams and the splitting power fraction is $0.8$, we can use the decoding order discussed in Section \ref{sec:3}-C and then calculate the rate for URLLC user 1 and URLLC user 2, and the results are $2.62\ \mathrm{(bits/s/Hz)}$ and $1.68\ \mathrm{(bits/s/Hz)}$, respectively. Now $r_U$ can be $1.68\ \mathrm{(bits/s/Hz)}$, so $r_U^{sum}$ also increases.

In Fig. \ref{fig:embb_urllc_geq}, $\Gamma_B=20$ \ dB, $\Gamma_U=10$ \ dB, $\epsilon_B=10^{-3}$, $\epsilon_U=10^{-5}$. This simulation result shows that OMA outperforms NOMA except when $r^B_{sum}$ is relatively high. RSMA can outperform NOMA for the same reason in Fig. \ref{fig:embb_urllc_leq}, and outperform OMA when $r^{sum}_B$ is in the range from $0-5 \ \mathrm{(bits/s/Hz)}$ and in the range from $25-40 \ \mathrm{(bits/s/Hz)}$. When $\Gamma_B>\Gamma_U$, it is more difficult to satisfy the reliability requirements of URLLC devices when interference from the eMBB device exists, so OMA is preferable in this scenario. However, if the aim is to achieve a relatively high $r_B^{sum}$, NOMA can perform better, because eMBB device can take advantage of frequency diversity. Similarly, RSMA can outperform OMA, and the improvement made by RSMA is more obvious when interference from eMBB device is relatively low, and this is also shown in Fig. \ref{fig:embb_urllc_leq}.

\begin{figure}[]
\centering
\includegraphics[scale=0.6]{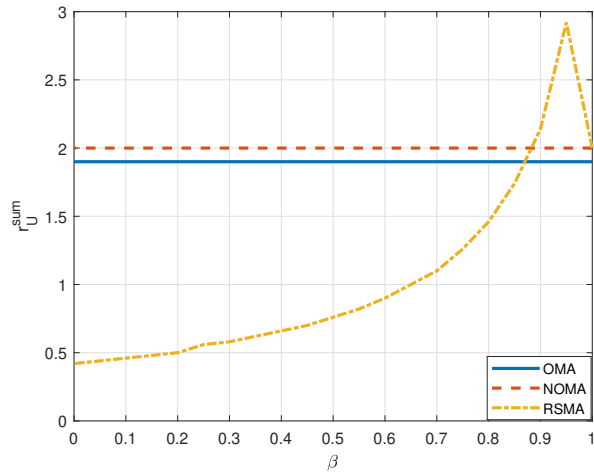}
\caption{$r_U^{sum}$ versus power factor $\beta$. $\Gamma_B=10$\ dB, $\Gamma_U=20$\ dB, $F=10$, $\epsilon_B=10^{-3}$, $\epsilon_U=10^{-5}$.}\label{fig:beta_B_10}
\end{figure}

\begin{figure}[]
\centering
\includegraphics[scale=0.6]{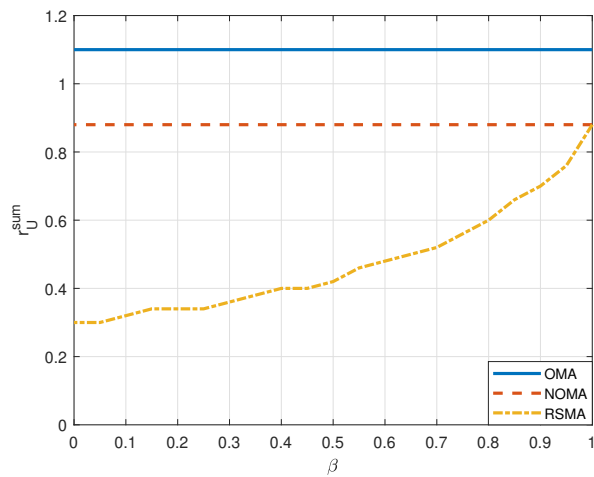}
\caption{$r_U^{sum}$ versus power factor $\beta$. $\Gamma_B=20$\ dB, $\Gamma_U=10$\ dB, $\epsilon_B=10^{-3}$, $\epsilon_U=10^{-5}$.}\label{fig:beta_B_20}
\end{figure}

Fig. \ref{fig:beta_B_10} and Fig. \ref{fig:beta_B_20} present the relation between splitting power fraction $\beta$ and the sum-rate of URLLC $r_U^{sum}$. URLLC users occupy $F_U=5$ frequency resources in the OMA case, and the eMBB user occupies another $F-F_U=5$ frequency resources. In both NOMA and RSMA cases, the eMBB rate is fixed at the rate which is the same as the one in OMA case. Since there is no power splitting in OMA and NOMA, the lines corresponding to OMA and NOMA are horizontal lines. In Fig. \ref{fig:beta_B_10}, $\Gamma_B=10$\ dB, $\Gamma_U=20$\ dB, and $r_U^{sum}$ achieves the maximal point at $\beta=0.95$. In this scenario, the URLLC channel condition is better than the eMBB one, so RSMA has more space to adjust the splitting power fraction. However, in Fig. \ref{fig:beta_B_20}, $\Gamma_B=20$\ dB, $\Gamma_U=10$\ dB, $r_U^{sum}$ increases as $\beta$ increases, and it achieves the maximum point at $\beta=1$, which means RSMA boils down to NOMA. The reason is that due to the relatively high eMBB interference, it is difficult for URLLC user 2 to achieve $r_U$ while having the interference from a split stream of URLLC user 1, so splitting power does not make that much difference in this situation. This explains why RSMA and NOMA have the same performance when the rate of eMBB is relatively high in Fig. \ref{fig:embb_urllc_geq}.

\begin{figure}
\centering
\begin{subfigure}[b]{0.5\textwidth}
   \includegraphics[scale=0.6]{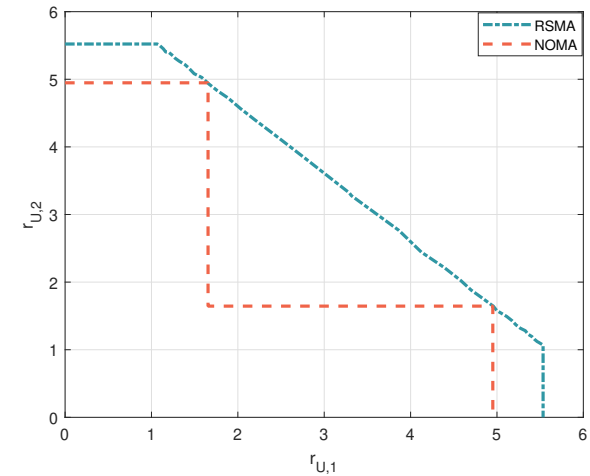}
   \caption{}
   \label{fig:Ng1} 
\end{subfigure}

\begin{subfigure}[b]{0.5\textwidth}
   \includegraphics[scale=0.6]{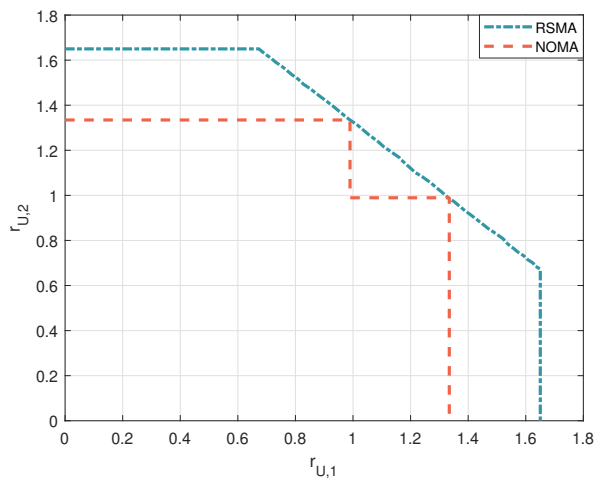}
   \caption{}
   \label{fig:Ng2}
\end{subfigure}

\caption[Two illustrations of rate region]{The rate regions $(r_{U,1}, r_{U,2})$ of two URLLC users in eMBB and URLLC coexistence scenario. (a) is the rate region when $\Gamma_B=10$\ dB, $\Gamma_U=20$ \ dB, $\epsilon_B=10^{-3}$ and $\epsilon_U=10^{-5}$. (b) is the rate region when $\Gamma_B=20$\ dB, $\Gamma_U=10$ \ dB, $\epsilon_B=10^{-3}$ and $\epsilon_U=10^{-5}$.}
\label{rate_region}
\end{figure}

The reason why RSMA outperforms NOMA is that it can provide a more flexible SIC decoding order to achieve all the boundary points of the rate region theoretically. Thus, a higher $r_U$ and $r_U^{sum}$ could be found in RSMA. In eMBB and URLLC coexistence, the rate regions of two URLLC users in RSMA and NOMA are shown in Fig. \ref{rate_region}, and the eMBB rate is the same as the rate in Fig. \ref{fig:beta_B_10} and Fig. \ref{fig:beta_B_20}. In Fig. \ref{rate_region}, the boundary points are the average of achievable rate pairs which satisfy the reliability requirements. NOMA without time-sharing and joint encoding/decoding can only achieve the points in the region bounded by the red line, while RSMA can achieve the points in the region bounded by the blue line, which includes the region of NOMA. The essence is that SINR of both URLLC users is adjustable by changing the splitting power fraction, so that RSMA can achieve all the points on the diagonal line instead of just two corner points in the rate region, and this can be seen as the user with higher rate compensates the user with lower rate to achieve the desired rate. When $\Gamma_B=10$\ dB and $\Gamma_U=20$\ dB, the SINR of URLLC users is high, the two corner points of NOMA are distant and the difference of regions of NOMA and RSMA is big, which is shown in Fig. \ref{fig:Ng1}. In this situation, RSMA has more room to adjust the splitting power fraction to obtain the points on the diagonal line, and a higher $r_U$ can be found. However, when $\Gamma_B=20$\ dB and $\Gamma_U=10$\ dB and the SINR of URLLC users is low, the two corner points of NOMA can be very close due to the high interference from eMBB, so the difference between NOMA and RSMA regions becomes smaller, which is shown in Fig. \ref{fig:Ng2}. In this scenario, RSMA only has a small room to play its role, and the user with a higher rate also struggles to satisfy the rate requirement, so it is more difficult to compensate the user with a lower rate. Thus, RSMA obtains better performance than both OMA and NOMA when $r_B^{sum}$ is relatively low, while obtains the same performance as NOMA when $r_B^{sum}$ is high.

\subsection{eMBB and mMTC Coexistence}

\begin{figure}[]
\centering
\includegraphics[scale=0.6]{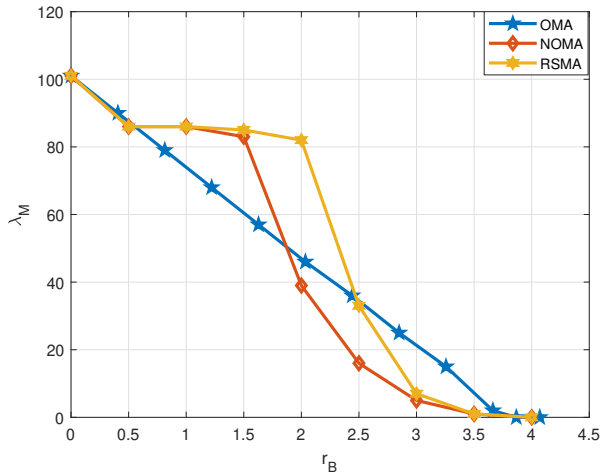}
\caption{Achievable pairs $(r_B, \lambda_M)$, where $r_B$, $\lambda_M$ are the rate of eMBB and the arrival rate of mMTC traffic, respectively. $\Gamma_B=20$\ dB, $\Gamma_M=5$\ dB, $r_M=0.04 \ \mathrm{(bits/s/Hz)}$, $\epsilon_B=10^{-3}$, $\epsilon_M=10^{-1}$.}\label{fig:embb_mmtc_20}
\end{figure}

The simulation results of achievable pairs $(r_B, \lambda_M)$ are presented. In Fig. \ref{fig:embb_mmtc_20}, $\Gamma_B=20$\ dB, $\Gamma_M=5$\ dB, $r_M=0.04 \ \mathrm{(bits/s/Hz)}$, $\epsilon_B=10^{-3}$, $\epsilon_M=10^{-1}$. This result shows that when $r_B>1.8 \ \mathrm{(bits/s/Hz)}$, OMA can achieve higher $\lambda_M$ than NOMA; while when $0.5<r_B<1.8$ $\mathrm{(bits/s/Hz)}$, NOMA achieves higher $\lambda_M$. RSMA always outperforms NOMA and outperforms OMA when $0.5<r_B<2.5 \ \mathrm{(bits/s/Hz)}$, but it still cannot achieve the region that OMA achieves when $r_B>2.5 \ \mathrm{(bit/s/Hz)}$. The curve of OMA is approximately linear, while for NOMA and RSMA the curves drop dramatically at $r_B=1.5 \ \mathrm{(bit/s/Hz)}$ and $r_B=2 \ \mathrm{(bit/s/Hz)}$, respectively.

When $r_B$ is relatively low, i.e. $0-0.5 \ \mathrm{(bits/s/Hz)}$, the performances of these three schemes are very similar, because for NOMA and RSMA the interference from eMBB device is relatively low, so it almost does not affect mMTC traffic, and for OMA, a large time fraction is allocated to mMTC traffic and eMBB only occupies a small one. When $r_B$ is medium, i.e. $0.5-2\ \mathrm{(bits/s/Hz)}$, RSMA and NOMA outperform OMA. At this eMBB rate, BS can leverage the differences between channel gains and reliability requirements of eMBB and mMTC devices when NOMA or RSMA is applied, but OMA cannot leverage these differences. When eMBB rate is relatively high, i.e. $2-4\ \mathrm{(bits/s/Hz)}$, OMA outperforms NOMA, and it also outperforms RSMA when $r_B>2.5\ \mathrm{(bits/s/Hz)}$. eMBB device will cause high interference to mMTC devices due to its relatively high rate, and almost no mMTC devices can be decoded before the eMBB device. Because of the power constraint of the eMBB device, the eMBB device cannot tolerate much interference from mMTC devices, so $\lambda_M$ drops to satisfy $\epsilon_M$. While for OMA, eMBB and mMTC traffic are isolated, so mMTC will not be affected by eMBB so much compared to NOMA and RSMA.

RSMA mainly outperforms NOMA at a medium eMBB rate, i.e. $1.5-3 \ \mathrm{(bits/s/Hz)}$. At this rate region, some mMTC devices must be decoded after eMBB due to the insufficient SINR, so if eMBB cannot be decoded, the decoding procedure is 'lodged' and then terminates. As analysed in the previous section, using RSMA can let part of the eMBB interference be cancelled, and in this way, the decoding procedure avoids 'lodged' to some extent, so RSMA has gain in this situation. For low eMBB rate and high eMBB rate, RSMA does not make much difference, because at a low rate almost all mMTC devices can be decoded before the eMBB device, and at a high rate mMTC devices must be decoded after eMBB device, and RSMA boils down to NOMA.

\begin{figure}[]
\centering
\includegraphics[scale=0.6]{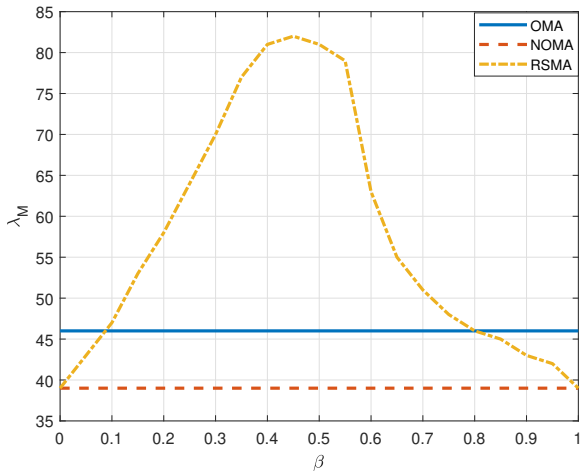}
\caption{Arrival rate $\lambda_M$ versus power factor $\beta$. $\Gamma_B=20$\ dB, $\Gamma_M=5$\ dB, $r_M=0.04 \ \mathrm{(bits/s/Hz)}$, $\epsilon_B=10^{-3}$, $\epsilon_M=10^{-1}$.}\label{fig:power_factor_beta}
\end{figure}

Fig. \ref{fig:power_factor_beta} shows the relation between mMTC arrival rate $\lambda_M$ and power factor $\beta$ when eMBB rate is fixed at $2  \ \mathrm{(bits/s/Hz)}$. Since OMA and NOMA do not split power, the corresponding curves are horizontal. For RSMA, the arrival rate achieves the maximum point at $\beta=0.45$, and is the same as NOMA when $\beta=0$ and $\beta=1$. As shown in Fig. \ref{fig:embb_mmtc_20}, $r_B=2 \ \mathrm{(bits/s/Hz)}$ is a relatively high rate, and either mMTC devices or eMBB device may not have enough SINR to be decoded, which causes decoding procedure to terminate. By adjusting the splitting power fraction of the eMBB device, the SINR of mMTC devices is also adjusted, so that more mMTC devices can be decoded. Thus, $\lambda_M$ increases as $\beta$ increases till $\beta=0.45$, and then decreases to the point which is as same as NOMA at $\beta=1$.

\section{Conclusion and Future Work}\label{sec:6}
In this paper, we investigate eMBB and URLLC coexistence and eMBB and mMTC coexistence with OMA, NOMA and RSMA. For RSMA scheme designs, it is important to decide properly which user splits the message and design the power allocations. Given the different requirements of eMBB, URLLC and mMTC, different splitting strategies are used. For eMBB and URLLC coexistence, RSMA is applied between URLLC users to fulfil the requirements of eMBB and URLLC. The performances in different channel conditions are simulated and analysed, and then we discuss the suitable scenario for RSMA. For eMBB and mMTC coexistence, a novel RSMA scheme is proposed. RSMA is applied between eMBB and mMTC because of service requirements and the nature of mMTC arrival process, and the performance is simulated and analysed. 

According to the simulation results, RSMA always outperforms NOMA in terms of achievable pairs for both coexistence scenarios. RSMA and NOMA can achieve the achievable pairs that OMA cannot; while OMA can also achieve the pairs that RSMA or NOMA cannot. The advantages of RSMA are that it can leverage the differences between channel gains and reliability requirements, and it achieves a larger rate region. The splitting power fraction of RSMA is a very significant factor to achieve the optimal sum-rate, and this paper gives insights on how the splitting power fraction can impact the performances. For eMBB and URLLC coexistence, the lower the interference from eMBB, the higher the improvement RSMA can make, because RSMA has more room to adjust the power allocation and obtain the optimal rate pairs while guaranteeing the service requirements. For eMBB and mMTC coexistence, due to splitting the messages, it is possible for RSMA to let part of the eMBB interference be cancelled, so that it can support a higher arrival rate of mMTC for a given reliability requirement when the splitting power fraction is set properly.  

This paper presents the promising results of RSMA with network slicing. These results will motivate us to explore more general setups. For eMBB and URLLC coexistence scenario, we do not consider finite blocklength, because in this work we consider that URLLC does not have instantaneous CSI, such that the factor which causes packet error is the randomness of the channel coefficient (SNR) instead of the noise instance at a given SNR \cite{7529226}. Therefore, we are working with the asymptotic rate. The scenario related to finite blocklength could be explored in future work. A more general $n_U$-URLLC-user scenario would also be an interesting direction and the extension of Remark 1 about decoding order could be studied.

\section*{Acknowledgments}
The work of Petar Popovski was supported in part by the Villum Investigator Grant “WATER” from the Velux Foundations, Denmark.

\bibliographystyle{IEEEtran} 
\bibliography{references.bib} 




%

\begin{IEEEbiography}[{\includegraphics[width=1in,height=1.25in,clip,keepaspectratio]{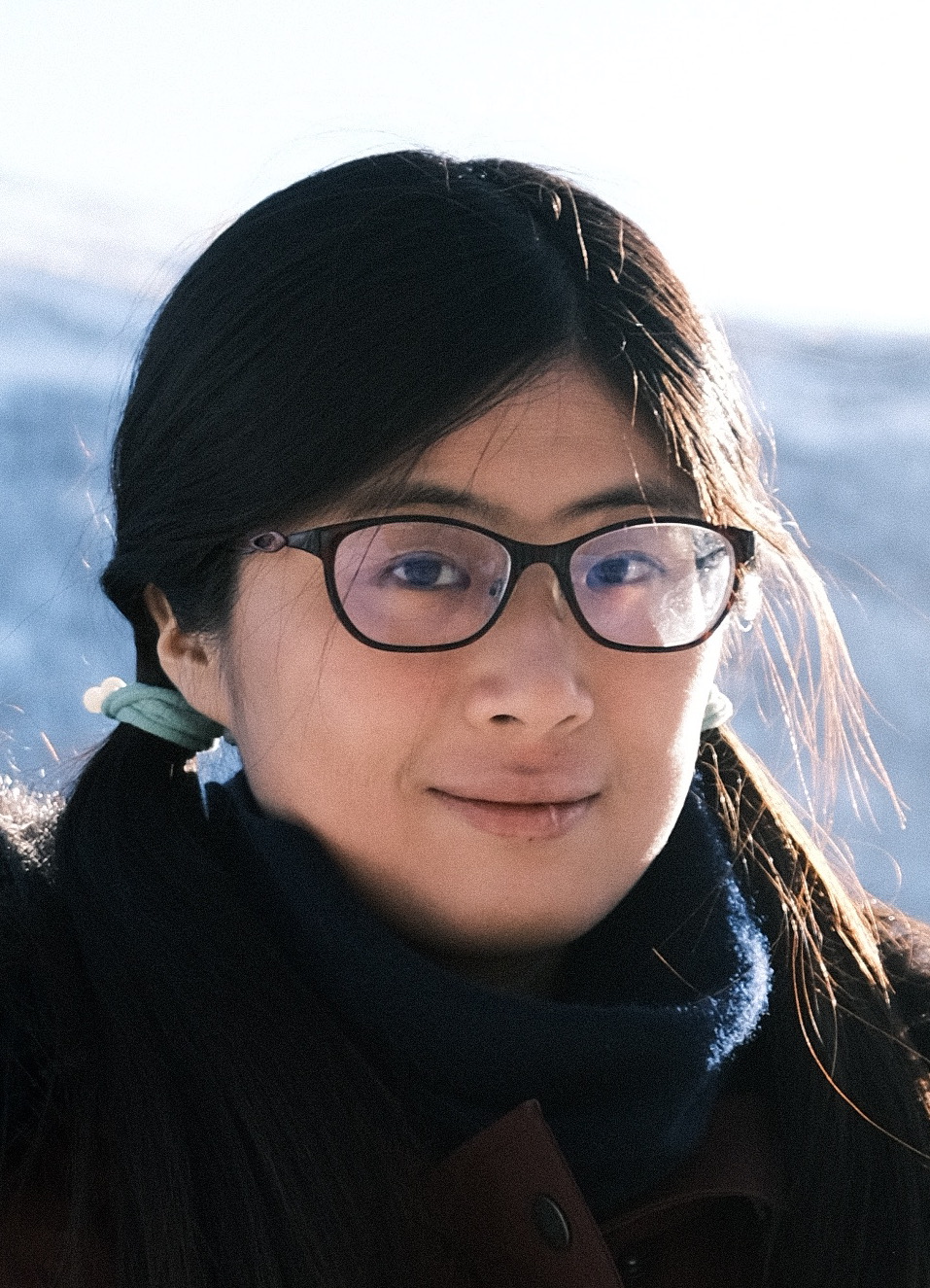}}]{Yuanwen Liu} Yuanwen Liu received the BEng degree in telecommunications engineering and M.S. in electrical engineering from Beijing University of Posts and Telecommunications Beijing, China in 2019 and Northwestern University, Evanston, United States in 2021. Currently she is a PhD student with the Department of Electrical and Electronic Engineering at Imperial College London, United Kingdom. Her research interests is rate splitting multiple access. 
\end{IEEEbiography}
\vfill
\begin{IEEEbiography}[{\includegraphics[width=1in,height=1.25in,clip,keepaspectratio]{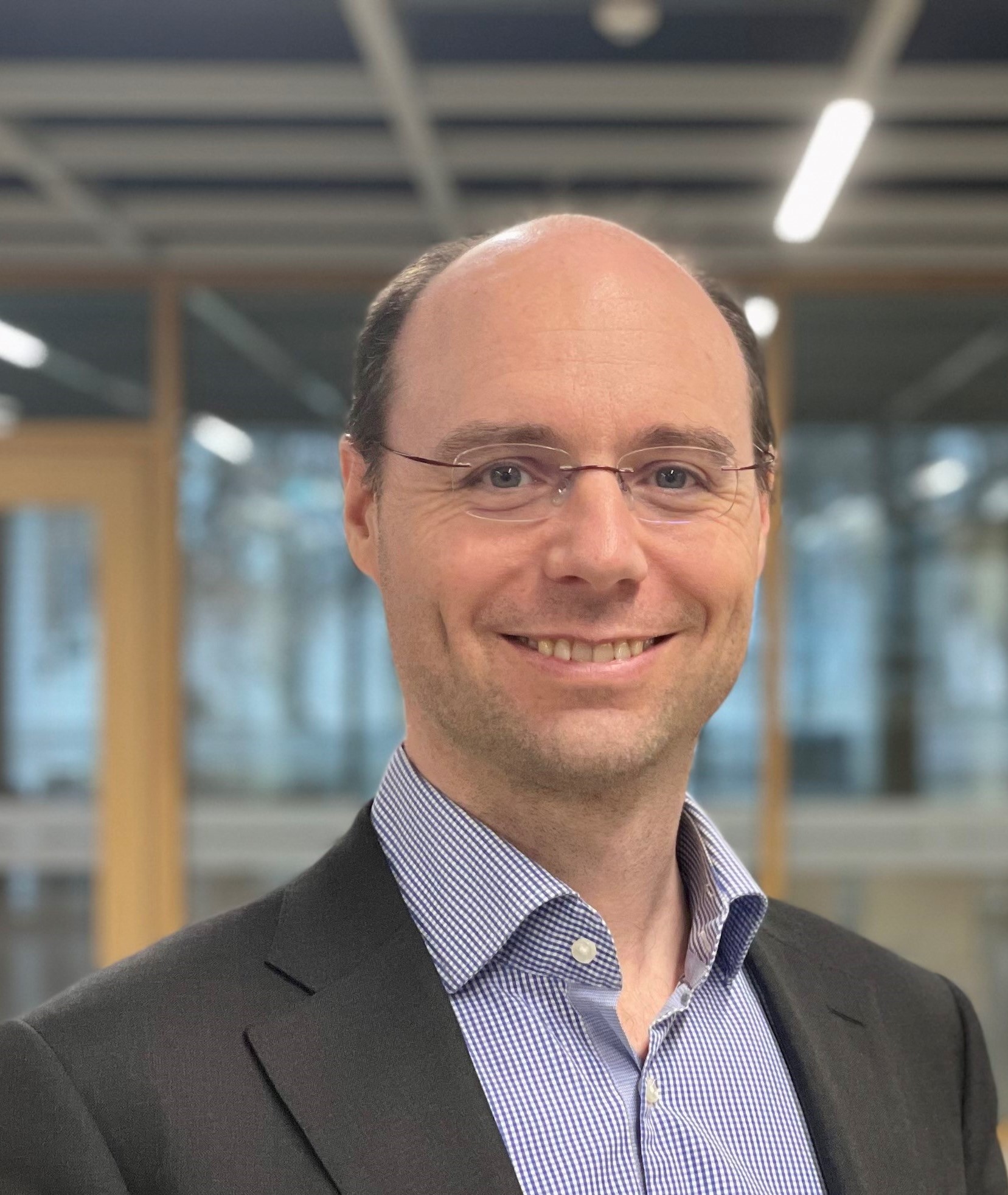}}]{Bruno Clerckx}
(Fellow, IEEE) Bruno Clerckx is a (Full) Professor, the Head of the Wireless Communications and Signal Processing Lab, and the Deputy Head of the Communications and Signal Processing Group, within the Electrical and Electronic Engineering Department, Imperial College London, London, U.K. He is also the Chief Technology Officer (CTO) of Silicon Austria Labs (SAL) where he is responsible for all research areas of Austria's top research center for electronic based systems. He received the MSc and Ph.D. degrees in Electrical Engineering from Université Catholique de Louvain, Belgium, and the Doctor of Science (DSc) degree from Imperial College London, U.K. Prior to joining Imperial College in 2011, he was with Samsung Electronics, Suwon, South Korea, where he actively contributed to 4G (3GPP LTE/LTE-A and IEEE 802.16m).
He has authored two books on “MIMO Wireless Communications” and “MIMO Wireless Networks”, 250 peer-reviewed international research papers, and 150 standards contributions, and is the inventor of 80 issued or pending patents among which several have been adopted in the specifications of 4G standards and are used by billions of devices worldwide. His research spans the general area of wireless communications and signal processing for wireless networks. He received the prestigious Blondel Medal 2021 from France for exceptional work contributing to the progress of Science and Electrical and Electronic Industries, the 2021 Adolphe Wetrems Prize in mathematical and physical sciences from Royal Academy of Belgium, multiple awards from Samsung, IEEE best student paper award, and the EURASIP (European Association for Signal Processing) best paper award 2022. He is a Fellow of the IEEE and the IET, and an IEEE Communications Society Distinguished Lecturer.
\end{IEEEbiography}

\begin{IEEEbiography}[{\includegraphics[width=1in,height=1.25in,clip,keepaspectratio]{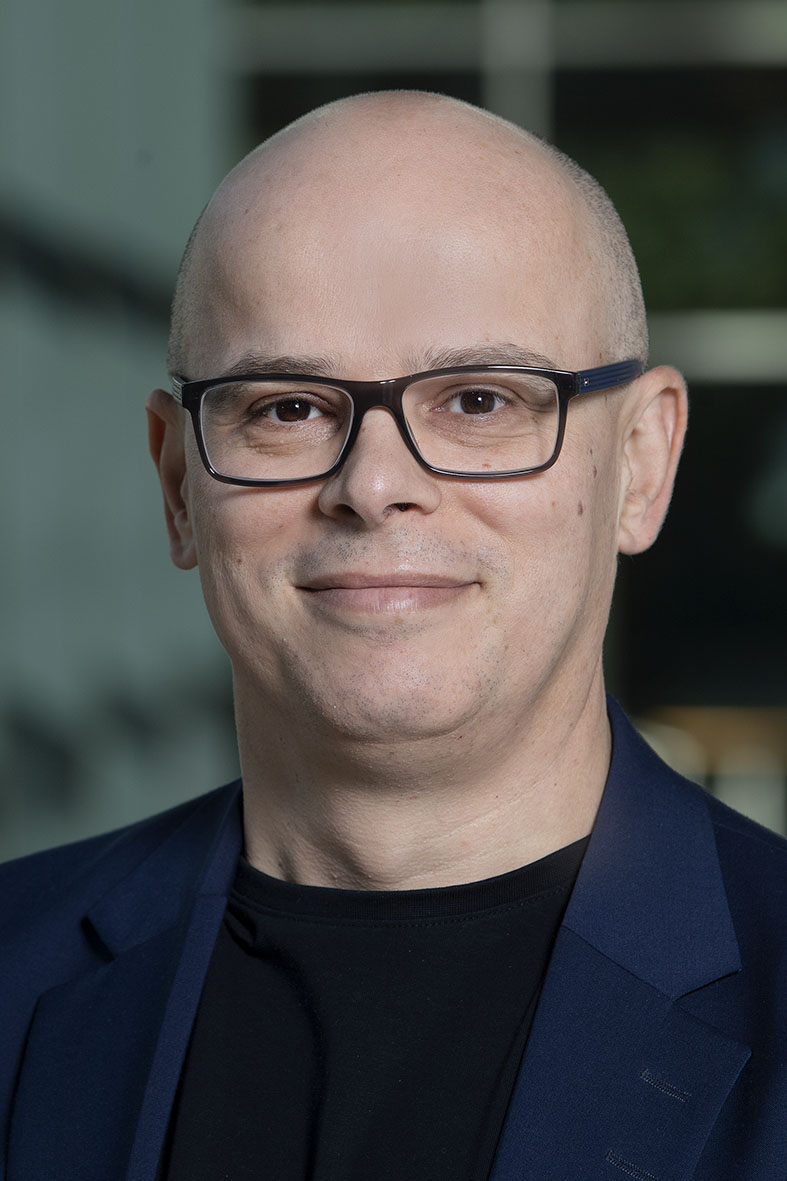}}]{Petar Popovski}
(Fellow, IEEE) is a Professor at Aalborg University, where he heads the section on Connectivity and a Visiting Excellence Chair at the University of Bremen. He received his Dipl.-Ing and M. Sc. degrees in communication engineering from the University of Sts. Cyril and Methodius in Skopje and the Ph.D. degree from Aalborg University in 2005. He received an ERC Consolidator Grant (2015), the Danish Elite Researcher award (2016), IEEE Fred W. Ellersick prize (2016), IEEE Stephen O. Rice prize (2018), Technical Achievement Award from the IEEE Technical Committee on Smart Grid Communications (2019), the Danish Telecommunication Prize (2020) and Villum Investigator Grant (2021). He was a Member at Large at the Board of Governors in IEEE Communication Society 2019-2021. He is currently an Editor-in-Chief of IEEE JOURNAL ON SELECTED AREAS IN COMMUNICATIONS and a Chair of the IEEE Communication Theory Technical Committee. Prof. Popovski was the General Chair for IEEE SmartGridComm 2018 and IEEE Communication Theory Workshop 2019. His research interests are in the area of wireless communication and communication theory. He authored the book ``Wireless Connectivity: An Intuitive and Fundamental Guide''.
\end{IEEEbiography}

\vfill






\end{document}